\documentclass[12pt]{article}
\usepackage{amsmath}
\usepackage{amssymb}
\usepackage{cite}
\newsymbol \blackbox 1004
\newcommand{\eh}{\hfill}\newlength{\sperr}

\newenvironment{proof}{{\settowidth{\sperr}{\bf\rm
Proof}%
\par\addvspace{0.3cm}\noindent\parbox[t]{1.3\sperr}
{\bf\rm P\eh r\eh o\eh o\eh f\eh }%
}}{\nopagebreak\mbox{}
$\blackbox$\par\addvspace{0.3cm}}

\def\a{\alpha}

\def\g{\gamma}

\def\d{\delta}

\def\vk{\varkappa}

\def\s{\sigma}
\def\la{\lambda}
\def\om{\omega}
\def\Om{\Omega}

\def\t{\theta}

\def\vp{\varphi}
\def\ve{\varepsilon}
\def\wh{\widehat}
\def\wt{\widetilde}
\def\ov{\overline}
\def\p{\partial}

\def\BC{{\mathbb C}}
\def\BR{{\mathbb R}}

\def\diag{\mathrm{diag}}
\newtheorem{Pa}{Paper}[section]
\newtheorem{Tm}[Pa]{{\bf Theorem}}

\newtheorem{Cy}[Pa]{{\bf Corollary}}
\newtheorem{Rk}[Pa]{{\bf Remark}}

\newtheorem{Ee}[Pa]{{\bf Example}}
\newtheorem{Dn}[Pa]{{\bf Definition}}
\newtheorem{Pn}[Pa]{{\bf Proposition}}

\title{Sine--Gordon theory in a semi--strip}

\author{Alexander Sakhnovich}

\date{}
\begin{document}
\maketitle

\noindent Address:

\noindent  {\it Faculty of Mathematics
 
\noindent  University of Vienna
 
\noindent  Nordbergstrasse 15
 
 \noindent A-1090 Vienna, Austria }

\noindent E-mail: {\it Oleksandr.Sakhnovych@univie.ac.at 
 }

\vspace{0.5cm}

\begin{abstract} 
Initial-boundary value problems for  complex sine-Gordon and
sine-Gordon equations in a semi--strip are treated.
The evolution of the Weyl function and a uniqueness result 
 are obtained for complex sine-Gordon equation. The evolution of 
 the Weyl function
 as well as an existence result and a procedure
 to recover solution are given for sine-Gordon equation.
It is shown that for a wide class of examples the solutions of the
sine-Gordon equation are unbounded in the quarter-plane.
\end{abstract}

\vspace{0.5cm}
\noindent 
{MSC(2010) 35G31, 37K15,  34A55, 34B20, 34L40.} \\

\noindent 
Keywords: \\ {\it sine-Gordon equation, complex sine-Gordon equation,\\
skew-self-adjoint Dirac system, Weyl function, \\  inverse problem,  initial-boundary
value problem.}

\vspace{0.5cm}

\section{Introduction} \label{intro}
\setcounter{equation}{0}
The well-known sine-Gordon and complex sine-Gordon equations
are actively used in study of various physical models and processes:
self-induced transparency and coherent optical pulse propagation,
relativistic vortices in a superfluid, nonlinear sigma models,
the motion of rigid pendula, dislocations in cristals and so on (see, for instance,
references in \cite{AS, Ba, BP, IR, PS2, Pol, SCM}).

{\it Sine--Gordon equation} (SGE) in the light cone coordinates has the form
\begin{equation}      \label{0.1}
\psi_{\xi  \eta}=2\sin 2\psi, \quad \psi_{\xi}:=\frac{\p}{\p \xi}\psi.
\end{equation}
It is the first equation  to which the so called auto-B\"acklund transformation was
applied. It is also a one of the first equations, for which a Lax pair was found
and which was  consequently solved by the  Inverse Scattering Transform method
\cite{AKNS1}. The  more general {\it complex sine--Gordon equation}
(CSGE) was introduced (and its integrability was treated) only several
years later \cite{LR, Pol}.  For further developments of the theory of CSGE and various
applications see, for instance,  \cite{Ba, BP, BoTz, DH, PS1, PS2} and references
therein. CSGE has the form
\begin{equation}      \label{0.2}
\psi_{\xi  \eta}+\frac{4 \cos \psi}{(\sin \psi)^3}\chi_{\xi  }\chi_{\eta}=2\sin 2\psi, 
\quad \chi_{\xi  \eta}-\frac{2}{\sin \, 2 \psi}\big(\psi_{\xi}\chi_{\eta}+\psi_{\eta}\chi_{\xi}\big)=0,
\end{equation}
where $\psi=\ov\psi$, $\chi=\ov\chi$, and $\ov\psi$ denotes the complex conjugate of  $\psi$.

There are also two constraint equations
\begin{equation}      \label{0.3}
2( \cos \psi)^2\chi_{\xi }-( \sin \psi)^2\t_{\xi }=2( \sin \psi)^2\g, \quad
2( \cos \psi)^2\chi_{\eta }+( \sin \psi)^2\t_{\eta }=0, 
\end{equation}
where $\g$ is a constant $( \g=\ov \g \equiv {\mathrm{const}})$
and $\t=\ov \t$.

For the particular
case $\chi \equiv 0$ and $\t=-2\xi\g$, CSGE turns into sine--Gordon equation \eqref{0.1}
and the constraint equations hold automatically. There are also many interesting
modifications and generalizations of the sine--Gordon equation: elliptic SGE, matrix SGE,
non-abelian SGE, sh-Gordon equation, et cetera, which often enough could be studied
in a way similar to SGE.

The introduction of the Inverse Scattering Transform  brought a breakthrough
in the initial value problems for  integrable nonlinear equations. The initial-boundary value
problems are more complicated, though the Inverse Scattering Transform
and several other methods help to obtain various interesting results.
In spite of many interesting developments the rigorous results in this domain
are comparatively rare.  Nevertheless one could mention, for instance, important
uniqueness and existence results in \cite{Abd, BSZ, Bona, CB, EY, Hol, Ton}  (see also references therein).
Here, we  apply the Inverse Spectral Transform method \cite{Ber, BerG, KvM, SaA1, SaA2, 
SaA4, SaA08, Sa88, SaL2,
SaL30, SaL3}.
In this way we shall obtain a uniqueness result for
CSGE and a global existence result for SGE in the semi-strip
\begin{equation}      \label{0.9}
{\cal D}=\{(\xi,\, \eta):\,0 \leq \xi <\infty, \,\,0\leq \eta<a\}.
\end{equation}
Notice that the initial-boundary problem for SGE, where
the values of $\psi$ are given on the characteristics
$\xi=-\infty$ and $\eta=0$, is treated in \cite{KN}
(see \cite{ZTF} for the related Cauchy problem for
SGE in laboratory coordinates). A local solution of the
Goursat problem for SGE, where $\psi$ is given on the 
characteristics $\xi=0$ and $\eta=0$, is described in
\cite{Kri} (see also \cite{LS}). Our result for SGE
is based on a global existence theorem from \cite{SaA2}.

Preliminaries on zero curvature equations for CSGE and SGE and on  inverse spectral 
problem are contained in the next Section \ref{Prel}.
Evolution of the Weyl function and uniqueness result for CSGE are given in Section  \ref{CSGE}, and existence
and recovery of solution of the initial-boundary value problem for SGE
are treated in Section \ref{SGE}. In the second subsection of Section 4
we construct a class of unbounded in the quarter-plane solutions of
the initial-boundary value problem.

As usual, we denote the real part of a scalar or matrix $z$ by $\Re z$ and the imaginary part
by $\Im z$. The real axis is denoted by $\BR$, the positive semi-axis by $\BR_+$, 
the complex plane is denoted by $\BC$, and the lower semi-plane - by $\BC_-$.

\section{Preliminaries} \label{Prel}
\setcounter{equation}{0}
\subsection{Zero curvature equations}
Zero curvature representation of the integrable nonlinear equations
is a well known approach (see \cite{AKNS, TF, Nov, ZM} and references therein), 
which was developed
soon after the seminal Lax pairs appeared in \cite{Lax}. We shall need
zero curvature equations for CSGE and SGE.

If  $\sin \, 2 \psi \not=0$
the compatibility condition for constraint equations \eqref{0.3} is equivalent to
the second equation in  \eqref{0.2}. If $\sin 2\psi\not=0$ and \eqref{0.3} is true,
then CSGE  (i.e.,  equations \eqref{0.2}) is equivalent to the compatibility condition
of the systems
\begin{align}      \label{0.4}
&W_{\xi}=GW, \quad 
W_{\eta}=FW;\\ \label{0.5}
&G(\xi,\eta,\la):=i(\la - \g)j+V(\xi,\eta), \quad F(\xi,\eta,\la):=-\frac{i}{\la}g(\xi,\eta)^*jg(\xi,\eta).
\end{align}
Here
\begin{align}  \label{0.6}
& j = \s_3=\left[
\begin{array}{cc}
1& 0 \\ 0 & -1
\end{array}
\right], \hspace{1em}g(\xi,\eta) = D_1(\xi,\eta)\left[\begin{array}{cc}
\cos\psi& i \sin \psi \\
 i \sin \psi &\cos\psi
\end{array}\right]D_2(\xi,\eta),\\
& \label{0.7'}
D_1=\exp\big\{i\big(\chi +\frac{\t}{2}\big)j\big\}, \,\, D_2=\exp\big\{i\big(\chi -\frac{\t}{2}\big)j\big\},
\\ \label{0.7}
& V=-g^*g_{\xi}-i\g(g^*jg-j).
\end{align}
In other words, CSGE is equivalent to the zero curvature equation
\begin{equation}      \label{0.8}
G_{\eta}-F_{\xi}+[G,F]=0, \qquad [G,F]:=GF-FG.
\end{equation}
We shall consider CSGE in the semi-strip \eqref{0.9}.
According to \cite{PS2, Pol} the following statement is true.
\begin{Pn}\label{zc} Let  $\, \{\psi(\xi,\eta), \, \chi(\xi,\eta), \, \t(\xi,\eta)\,\}$ be a triple of real--valued 
and twice continuously differentiable functions on ${\cal D}$. Assume that $\sin 2\psi \not=0$
and that equations \eqref{0.2} and \eqref{0.3} hold. Then the zero curvature equation
\eqref{0.8} holds too.
\end{Pn}
Moreover,  $g$ given by \eqref{0.6} belongs $SU(2)$ and  satisfies relations
\begin{align}\label{0.10}
&g^*g_{\xi}+i\g g^*jg=i\big(\chi_{\xi}+\frac{1}{2}\t_{\xi}+\g\big)g^*jg+i\big(\chi_{\xi}-\frac{1}{2}\t_{\xi}\big)j
+i\psi_{\xi}D_2^*\s_1 D_2, \\
&\label{0.11}
g^*jg=D_2^* \left[
\begin{array}{lr}
\cos 2\psi&  i\sin 2 \psi \\ -i\sin 2 \psi &-\cos 2 \psi
\end{array}
\right]D_2, \quad
 \s_1:=\left[
\begin{array}{cc}
0& 1 \\ 1 & 0
\end{array}
\right].
\end{align}
In view of the first constraint in \eqref{0.3} and equalities \eqref{0.10} and \eqref{0.11},
the matrix function $V$ introduced by \eqref{0.7} has the form
\begin{equation}      \label{0.12}
V=\left[
\begin{array}{lr}
0& v \\ - \ov v & 0
\end{array}
\right], \quad v=\big(-i\psi_{\xi}+2\chi_{\xi}\cot\psi\big)e^{i(\t-2\chi)}.
\end{equation}
According to \eqref{0.5} and \eqref{0.12} the auxiliary system $W_x=GW$
is a Dirac-type system, which will be used for the study of CSGE.

In the zero curvature representation \eqref{0.8} of the sine-Gordon 
equation \eqref{0.1} we put
\begin{align}      \label{0.13}
& G(\xi, \eta,z)=izj+V(\xi,\eta), \quad v(\xi,\eta)=-\psi_{\xi}(\xi,\eta), 
\quad \psi=\ov \psi,
\\
& \label{0.14}
 V=\left[
\begin{array}{lr}
0& v \\ -  v & 0
\end{array}
\right], \quad F(\xi, \eta,z)=\frac{1}{i z}
\left[
\begin{array}{lr}
\cos 2\psi&  \sin 2 \psi \\ \sin 2 \psi &-\cos 2 \psi
\end{array}
\right],
\end{align}
though the pair $\wh G=\wh D  G\wh D^{-1}$, $\wh F=\wh D  F \wh D^{-1}$,
where $G$ and $F$ are defined in \eqref{0.13} and \eqref{0.14},
$\wh D:=\diag\{i,1\}$, and diag means diagonal matrix,
would be closer to $G$ and $F$ from \eqref{0.5}.

\subsection{Weyl function and inverse problem}
Put $z=\la -\g$ and write down the auxiliary system 
given by the first relations  in \eqref{0.4} and \eqref{0.5} in  the form
\begin{equation}      \label{1.1}
\frac{d}{d\xi}w(\xi, z)=G(\xi, z)w(\xi, z), \quad  G(\xi, z)=izj+V(\xi),   \quad w(0,z)=I_2,
\end{equation}
where $V$ has the form given by the first relations in \eqref{0.12}
and \eqref{0.14}, $I_2$ is the 
$2\times 2$ identity matrix, and $w$ is the normalized fundamental solution.
In this subsection we adduce some results on the Weyl theory of the skew-self-adjoint Dirac-type
system \eqref{1.1} from \cite{SaA1, SaA2} (see also \cite{CG2, AGKLS, GesST, KaS, MST, SaA3, SaA8, SaA08}).
\begin{Dn} \label{Dn1} 
Let system \eqref{1.1} be given on the semi-axis $[0, \, \infty)$.
Then a  function $\vp(z)$ holomorphic  in
some semi-plane $\Im z < -M<0$ is called a Weyl function of
this system, if 
\begin{equation}  \label{1.2}
\sup_{\xi \leq r, \, \Im z<-M} \left\|e^{i \xi z} w(\xi, z)
\left[\begin{array}{c}
\vp(z) \\ 1
\end{array}
\right]\right\|<\infty \quad {\mathrm{for}} \, {\mathrm{all}} \, \, 0<r< \infty .
\end{equation}
\end{Dn}
We shall consider systems  \eqref{1.1}, where potentials $v$ are  bounded  on all finite intervals:
\begin{equation} \label{1.2'}
\sup_{0<\xi<r}|v(\xi)| <\infty  \quad {\mathrm{for}} \, {\mathrm{all}} \, \, 0<r< \infty .
\end{equation}
The next statement follows from \cite{SaA3} (see Remark 8.4, p.113).
\begin{Pn}\label{P1} There is at most one Weyl function of system \\ \eqref{1.1},
where $v$ satisfies \eqref{1.2'}.
\end{Pn}
If $v$ is bounded on $[0, \, \infty)$ the Weyl function always exists.
\begin{Pn}\label{P2} \cite{SaA1, SaA2, SaA3}
If the inequality
\begin{equation} \label{1.3}
\sup_{0<\xi<\infty}|v(\xi)| \leq M
\end{equation}
holds, then there is a unique Weyl function $\vp$ of system \eqref{1.1}.
Moreover, this Weyl function is the unique function such that
\begin{equation} \label{1.4}
\int_0^\infty \left[ \begin{array}{lr}   \ov{ \varphi (z)} &
1
\end{array} \right]
  w(\xi, z)^*
w(\xi, z)
\left[ \begin{array}{c}
  \varphi (z) \\ 1 \end{array} \right] dx < \infty, \quad  \Im z<-M<0.
\end{equation}
The Weyl function is given by the formulas
\begin{align} \label{r1}&
\vp(z)=\lim_{r \to \infty}\frac{{\cal A}_{11}(r,z)P_1(r,z)+{\cal A}_{12}(r,z)P_2(r,z)}
{{\cal A}_{21}(r,z)P_1(r,z)+{\cal A}_{22}(r,z)P_2(r,z)} \qquad (\Im z<-M),\\ \label{r2} &
{\cal A}(r,z)=\big\{{\cal A}_{kp}(r,z)  \big\}_{k,p=1}^2:=w(r,\ov z)^*, 
\end{align}
where $P_1(r,z)$, $P_2(r,z)$ is an arbitrary non-singular pair of meromorphic
functions with property-$j$, namely
\begin{equation} \label{r3}
|P_2|^2>0, \qquad |P_2|^2 \geq |P_1|^2.
\end{equation}
Finally, we have  $|\vp(z)|\leq 1$ for $\Im z<-M$.
\end{Pn}
We shall need a scalar version (i.e., the case where $v(\xi)$ is scalar) of  Theorem 1 from \cite{SaA1}:
\begin{Tm} \label{TmInv} System \eqref{1.1} satisfying condition 
\eqref{1.3} is uniquely defined by its Weyl function.
\end{Tm}
The next theorem deals with the inverse problem for $\vp$
such that
\begin{equation} \label{1.4!}
\sup_{\Im z <-M}|z^2\big(\vp(z)-\a/z \big)| <\infty , \quad \a \in \BC.
\end{equation}
\begin{Tm} \label{TmInv2} \cite{SaA2, SaA3} Let holomorphic function
$\vp(z)$ $(\Im z<-M)$ admit representation \eqref{1.4!}.
Then there is a unique system \eqref{1.1} satisfying \eqref{1.2'} such that
$\vp$ is its Weyl fuction.
\end{Tm}
To recover $v(\xi)$  notice that according to \eqref{1.1} we have
\begin{align}\label{d1}
w(\xi, z)w(\xi, \ov z)^*=w(\xi, \ov z)^*w(\xi,  z)=I_2 ,
\end{align}
and
\begin{equation}\label{1.4!!}
v(\xi)=w_1^{\prime}(\xi)w_2(\xi)^*,   \,\, w_1(\xi):=[1 \quad 0]w(\xi,0),
\,\, w_2(\xi):=[0 \quad 1]w(\xi,0).
\end{equation}
To recover $w_1$ and $w_2$  on some interval $[0, \, c]$ ($c<\infty$) 
we construct function $s(\xi)\in L^2(0,c)$ via the Fourier transform
\begin{equation}\label{1.5}
\displaystyle{s(\xi)= \frac{i}{2 \pi}e^{- y \xi}{\mathrm{
l.i.m.}}_{b \to \infty} \int_{- b}^{b}e^{i \xi
x} (x +i y)^{-1} \vp\Big(\frac{x +i y}{ 2}\Big) d x, \quad  y
<-2M,}
\end{equation}
where l.i.m. is the limit in the norm of $L^2(0,c)$.
As $|\vp|\leq 1$ it is easy to see that $s$ does not depend on the choice
of  $y<-2M$.  
The function $s$ is  absolutely continuous and we have
\begin{equation}\label{1.6}
s(0)=0, \quad \sup_{0<\xi<c}|s^{\prime}(\xi)| <\infty, \quad s^{\prime}:=\frac{d}{d \xi}s
\end{equation}
(see (18) in \cite{SaA2}, where $\Phi_2=s$).
By  \cite{SaA2} the operator
\begin{equation}\label{1.7}
S_rf=f(\xi)+ \frac{1}{2}\int_0^r
\int^{\xi+u}_{|\xi-u|}s^{\prime} \Big(
\frac{l+\xi-u}{2} \Big) \ov{s^{\prime} \Big( \frac{l+u-\xi}{2} \Big)}d l f(u) du,
\end{equation}
where $0<r\leq c$,
is bounded in $L^2(0,r)$.  The inequality $S_r \geq I$, where $I$ is the identity
operator, holds.
Now, we can recover $w_1,\, w_2 \in\BC_2$ by the formulas
\begin{align}\label{1.8}
&\om_1(r)=-\int_0^r\ov{\big(S_r^{-1}s^{\prime}\big)(\xi)}d\xi, \quad 
\om_2(r)=1-\int_0^r\ov{\big(S_r^{-1}s^{\prime}\big)(\xi)}s(\xi)d\xi, \\
&\label{1.9}
w_2(r)=[\om_1(r) \quad \om_2(r)], \quad w_1(r)=[\ov{\om_2(r) }\quad - \ov{\om_1(r)}].
\end{align}
\begin{Tm}\label{TmR}
\cite{SaA1,SaA2,SaA3} Let function $\vp$ be the Weyl function of system
\eqref{1.1} such that \eqref{1.2'} holds. Assume that either \eqref{1.3} or \eqref{1.4!} is fulfilled.
Then the solution of the  inverse problem to recover $v$
is given by the formulas \eqref{1.4!!}, \eqref{1.8}, and \eqref{1.9},
where $s$ and $S_r$ are constructed  using formulas \eqref{1.5}  and \eqref{1.7}.   
\end{Tm}
\section{CSGE: evolution of the Weyl function and uniqueness of the solution} \label{CSGE}
\setcounter{equation}{0}
In this section we consider the initial-boundary value problem for CSGE:
\begin{equation}\label{2.1}
v(\xi,0)=h_1(\xi), \quad \psi(0,\eta)=h_2(\eta), \quad \chi(0,\eta)=h_3(\eta), \quad \t(0,0)=h_4,
\end{equation}
where $v$ is defined by the second equality in  \eqref{0.12}.
 We assume that the conditions of Proposition \ref{zc} are fulfilled.
\begin{Tm} \label{evol} Let  $\{\psi(\xi,\eta), \, \chi(\xi,\eta), \, \t(\xi,\eta)\}$ be a triple of real-valued 
and twice continuously differentiable functions on ${\cal D}$.
 Assume that $\sin 2\psi \not=0$, that $v$ is bounded, that is,
\begin{equation} \label{2.2}
\sup_{(\xi, \eta)\in {\cal D}}|\psi_{\xi}+2i\chi_{\xi}\cot\psi | \leq M,
\end{equation}
and that relations \eqref{0.2}, \eqref{0.3}, and \eqref{2.1} hold.
Then the Weyl functions $\vp(\eta,z)$ of the auxiliary Dirac-type systems
$\, W_{\xi}=GW$, where $\displaystyle{\,G(\xi,\eta,z)\, }$  $\displaystyle{(z=\lambda -\g)}$ is defined via \eqref{0.5} and \eqref{0.12},
exist and have the form
\begin{equation} \label{2.3}
\vp(\eta,z)=\frac{R_{11}(\eta,z)\vp(0,z)+R_{12}(\eta,z)}
{R_{21}(\eta,z)\vp(0,z)+R_{22}(\eta,z)}.
\end{equation}
Here $R:=\big\{R_{kp} \big\}_{k,p=1}^2$ is defined by the equalities
\begin{align}\nonumber 
\frac{d}{d \eta}R(\eta,z)= &\frac{1}{i (z+\g)}e^{-id(\eta)j}
\left[
\begin{array}{lr}
\cos 2h_2(\eta)& i \sin 2 h_2(\eta) \\ -i \sin 2 h_2(\eta) &-\cos 2 h_2(\eta)
\end{array}
\right] \\
&\label{2.4} \times
e^{id(\eta)j} R(\eta,z), 
\end{align}
\begin{equation}
 \label{2.4!}
 R(0,z)=I_2, \quad d(\eta)=h_3(0)-\frac{1}{2}h_4+\int_0^\eta h_3^{\prime}(u)\big(\sin h_2(u)\big)^{-2}du,
\end{equation}
and $\vp(0,z)$ is the Weyl function of the system
\begin{equation} \label{2.5}
\frac{d}{d \xi}W(\xi,z)=\big(izj+V(\xi)\big)W(\xi,z), \quad V(\xi)=
\left[
\begin{array}{lr}
0&h_1(\xi)\\-\ov{h_1(\xi)}&0
\end{array}
\right].
\end{equation}
\end{Tm}
\begin{proof}. Introduce normalized fundamental solutions of the auxiliary systems
by the the equalities:
\begin{align} \label{2.6}
&\frac{d}{d \xi}w(\xi,\eta,z)=G(\xi,\eta,z)w(\xi,\eta,z), \quad w(0,\eta,z)=I_2;
\\ \label{2.7'} &
\frac{d}{d \eta}R(\xi,\eta,z)=F(\xi,\eta,z)R(\xi,\eta,z), \quad R(\xi,0,z)=I_2; \\
& \label{2.7} 
R(\eta,z)=R(0,\eta,z);
\end{align}
where $G$ and $F$ are given by formulas \eqref{0.5},  \eqref{0.7'},  \eqref{0.11}, 
and  \eqref{0.12}. Notice that the definition of $R(\eta,z)$ in \eqref{2.7} complies
with \eqref{2.4} and  \eqref{2.4!}. The definition of $V$ in \eqref{2.5} complies with
\eqref{0.12} and \eqref{2.1}.
As $G$ and $F$ are continuously differentiable,
we can use factorization formula (1.6) from Chapter 12 in \cite{SaA3}:
\begin{equation} \label{2.8}
w(\xi,\eta,z)=R(\xi,\eta,z)w(\xi,0,z)R(\eta,z)^{-1}.
\end{equation}
From  \eqref{0.5} and \eqref{2.7'} we derive 
\[
\big(R(\xi,\eta,\ov z)^*R(\xi,\eta,z)\big)_{\eta}=0, \quad
{\mathrm{and}}\,\, {\mathrm{so}} \quad R(\xi,\eta,\ov z)^*=R(\xi,\eta,z)^{-1}, 
\]
$\det R(\xi,\eta,z)\not=0.$ In particular, we have
\begin{equation} \label{2.9}
\big(R(\eta,\ov z)^{-1}\big)^*=R(\eta,z).
\end{equation}
Taking into account  \eqref{2.8} and  \eqref{2.9} we get
\begin{equation} \label{2.10}
{\cal A}(\xi,\eta,z):=w(\xi,\eta,\ov z)^*=R(\eta,z){\cal A}(\xi,0,z)R(\xi,\eta,\ov z)^*.
\end{equation}
Now, notice that according to \eqref{2.2} the conditions of Proposition \ref{P2}
are fulfilled for any $a>\eta \geq 0$. 
It is immediate also that the pair 
$P_1=0$, $P_2=1$ is non-singular with property-$j$. 
Hence \eqref{r1} holds for these $P_1$ and $P_2$.
From   \eqref{r1} and \eqref{2.10} we get
\begin{equation} \label{2.10!}
\vp(\eta,z)=\lim_{r\to \infty}\frac{R_{11}(\eta,z)\wt P_1(r,\eta,z)+R_{12}(\eta,z)\wt P_2(r,\eta, z)}
{R_{21}(\eta,z))\wt P_1(r,\eta,z)+R_{22}(\eta,z))\wt P_2(r,\eta,z)},
\end{equation}
where
\begin{equation} \label{2.10!!}
\left[
\begin{array}{c}
\wt P_1(r,\eta,z) \\ \wt P_2(r,\eta,z)
\end{array}
\right]:={\cal A}(r,0,z)R(r,\eta,\ov z)^*\left[
\begin{array}{c}
0 \\ 1
\end{array}
\right].
\end{equation}

If $a<\infty$ and $|z+\g|>\d>0$ we have
\begin{equation} \label{2.11}
\sup |\eta/(z+\g)|=C<\infty,
\end{equation}
and so  the inequality
\begin{equation} \label{2.12}
\|R(\xi,\eta,z)-I_2-\int_0^\eta F(\xi,t,z)dt\| \leq \eta^2(z+\g)^{-2}e^{C}
\end{equation}
holds. Indeed, using \eqref{0.5}, \eqref{2.7'}, and the corresponding {\it multiplicative integral}
representation of $R$ we get
\begin{align} \label{2.13}
&R(\xi,\eta,z)=\lim_{n\to \infty}\prod_{k=1}^n\Big(I_2+\frac{\eta F_k}{n(z+\g)}+
\frac{\eta^2 C_k(n)}{n^2(z+\g)^2}\Big), 
\\  \label{2.14}& F_k:=-ig(\xi, \frac{k}{n}\eta )^*jg(\xi, \frac{k}{n}\eta ),
\quad \|C_k(n)\| \leq 1\quad{\mathrm{ for}} \quad  {\mathrm{all}} \quad  n>n_0(C).
\end{align}
It follows from \eqref{2.13} and  \eqref{2.14}  that for any $\ve>0$ and sufficiently large values of $n$
we have
\begin{align} \nonumber
& \|R(\xi,\eta,z)-I_2-\sum_{k=1}^n\frac{\eta F_k}{n(z+\g)}\| \leq
\Big(\frac{(1+\ve)\eta}{z+\g}\Big)^2
\\& \label{2.15} \times
\sum_{p=2}^n\Big(\frac{(1+\ve)\eta}{z+\g}\Big)^{p-2}\frac{1}{n^p}
\left( \begin{array}{c}n\\ p\end{array}\right)+\ve .
\end{align}
Inequalities \eqref{2.11} and \eqref{2.15} imply \eqref{2.12}.

According to \eqref{2.12} there is a value $M_1>M>0$ such that the pairs
\begin{equation} \label{2.16}
\left[
\begin{array}{c}
\wh P_1(r,\eta, z) \\ \wh P_2(r,\eta, z)
\end{array}
\right]:=R(r,\eta,\ov z)^*\left[
\begin{array}{c}
0 \\ 1
\end{array}
\right]
\end{equation}
are non-singular with property-$j$ for all $\eta<a$ and $z$ such that $\Im z<-M_1$.
Next, notice that in view of  \eqref{2.10!!} and \eqref{2.16} we get
\begin{equation} \label{2.17}
\left[
\begin{array}{c}
\wt P_1(r,\eta,z) \\ \wt P_2(r,\eta,z)
\end{array}
\right]:={\cal A}(r,0,z)\left[
\begin{array}{c}
\wh P_1(r,\eta,z) \\ \wh P_2(r,\eta,z)
\end{array}
\right].
\end{equation}
It is also easy to see that 
\begin{equation} \label{2.17'}
j>{\cal A}(r,\eta,z)^*j{\cal A}(r,\eta,z) \quad {\mathrm{for}}  \quad \Im z<-M_1, \quad r>0.
\end{equation}
Indeed, by \eqref{1.1} and \eqref{2.2} we have
\[
\frac{d}{d \xi}\big(w(\xi,\eta,z)^*jw(\xi,\eta,z)\big)> 0, \quad \Im z<-M,
\]
and so the inequality
\begin{equation} \label{2.17!}
 w(\xi,\eta,z)^*jw(\xi,\eta,z)> j \quad (\Im z<-M, \, \, \xi>0)
\end{equation}
is  true.  By \eqref{d1} we get $w(\xi,\eta,\ov{z})^*=w(\xi,\eta,z)^{-1}$.
Therefore, inequality \eqref{2.17'} is immediate from \eqref{r2} and   \eqref{2.17!}.

As the pair $\wh P_1$, $\wh P_2$ has property-$j$ and  \eqref{2.17'} holds,
it follows from \eqref{2.17} that
\begin{equation} \label{2.18}
\wt P_2(r,\eta,z) \not=0, \quad r>0.
\end{equation}
Moreover, according to \eqref{r1} and \eqref{2.17} we have
\begin{equation} \label{2.19}
\lim_{r\to \infty}\big(\wt P_1(r,\eta,z)/ \wt P_2(r,\eta,z)\big)=\vp(0,z).
\end{equation}
The evolution formula \eqref{2.3} is immediate from \eqref{2.10!} and \eqref{2.19}.

We assumed first that $a<\infty$ but if $a=\infty$ formula \eqref{2.3} is still proved for $\eta$ 
on all finite  intervals on $\BR_+$, and so \eqref{2.3} holds on $\BR_+$.
\end{proof}
The idea of the proof above as well as the idea to present the evolution
of the Weyl functions in the form of linear-fractional (M\"obius) transformations
comes from the seminal works \cite{SaL20, Sa88, SaL2}.
\begin{Cy}\label{Uniq} There is at most one triple $\{\psi(\xi,\eta), \, \chi(\xi,\eta), \, \t(\xi,\eta)\}$  of real-valued 
and twice continuously differentiable functions on ${\cal D}$ such that $\sin 2\psi \not=0$, that $v$ is bounded, and
that CSGE \eqref{0.2}, constraints \eqref{0.3}, and initial-boundary conditions \eqref{2.1} are satisfied.
\end{Cy}
\begin{proof}. Suppose the triple $\{\psi(\xi,\eta), \, \chi(\xi,\eta), \, \t(\xi,\eta)\}$
satisfies conditions of the corollary. Then the conditions of Theorem \ref{evol}
are fulfilled and the evolution of the Weyl function is given by formula \eqref{2.3}.
Hence, according to Theorem \ref{TmInv} the function $v(\xi,\eta)$ is uniquely recovered.

Next, we recover $\psi$ from $v$. Using \eqref{0.2}, \eqref{0.3}, and \eqref{0.12} we derive
\begin{align} \nonumber
&v_{\eta}=\big(-i\psi_{\xi \eta}+2\chi_{\xi \eta}\cot\psi-2\chi_{\xi}\psi_{\eta}(\sin\psi)^{-2}
\\
\nonumber
&
+
(\psi_{\xi}+2i\chi_{\xi }\cot\psi)(\t_{\eta}-2\chi_{\eta})\big)
 e^{i(\t-2\chi)}
=\big(-i\psi_{\xi \eta}+2\chi_{\xi \eta}\cot\psi
\\ \label{2.20} &
-2\chi_{\xi}\psi_{\eta}(\sin\psi)^{-2}
-2(\psi_{\xi}+2i\chi_{\xi }\cot\psi)
\chi_{\eta}(\sin\psi)^{-2}\big)e^{i(\t-2\chi)}
\\&\nonumber
=
\big(-2i\sin2\psi+2(\chi_{\xi \eta}-\frac{1}{\sin\psi \cos\psi}(\chi_{\xi}\psi_{\eta}+
\chi_{\eta}\psi_{\xi}))\cot\psi\big)e^{i(\t-2\chi)} \\
&=-2ie^{i(\t-2\chi)}\sin2\psi .
\nonumber \end{align}
It follows from \eqref{2.20} that $|\sin 2\psi|=|v_{\eta}|/2$. 
As  $\psi$ is continuous and $\sin 2\psi\not=0$,  the function $\sin 2 \psi$ is uniquely recovered from
the values of $|\sin 2\psi(\xi,\eta)|$ and the sign of $\sin 2 \psi(0,\eta)=\sin 2 h_2(\eta)$.
In view of \eqref{0.12} and \eqref{2.20} we get $\psi_{\xi}=2(\sin2\psi)\Re(v/v_{\eta})$,
and thus we recover also $\psi_{\xi}$. Therefore, the function
$\psi(\xi,\eta)=h_2(\eta)+\int_0^{\xi}\psi_x(x,\eta)dx$ is uniquely recovered too.
Moreover, we have 
\[
\chi(\xi,\eta)=h_3(\eta)+\int_0^{\xi}\chi_x(x,\eta)dx, \quad
\chi_{\xi}=2(\sin \psi)^2\Im (v/v_{\eta}).
\]
Finally,
$\t$ is uniquely recovered from the value $\t(0,0)$ and constraint equations
\eqref{1.3}.
\end{proof}
\begin{Rk}
 In a way similar to  the cases of other nonlinear equations 
 (see \cite{SaA2, SaL20, SaL3}) one can show that the evolution of the
Weyl function prescribed by CSGE  satisfies Riccati equation.  
Indeed, rewrite \eqref{2.3} as
\begin{align}& \label{2.22}
\vp(\eta,z)=\vk_1(\eta,z)/\vk_2(\eta,z), \quad \vk_1(\eta,z):=[1 \quad 0]R(\eta,z)
\left[
\begin{array}{c}
\vp(0,z) \\1
\end{array}
\right], \\
&\nonumber
\vk_2(\eta,z):=[0 \quad 1]R(\eta,z)
\left[
\begin{array}{c}
\vp(0,z) \\1
\end{array}
\right].
\end{align}
It is immediate from \eqref{2.7'}, \eqref{2.7}, and \eqref{2.22} that 
\begin{align}\nonumber &
\frac{d}{d \eta}\vp(\eta,z)=[1 \quad 0]F(0,\eta,z)R(\eta,z)
\left[
\begin{array}{c}
\vp(0,z) \\1
\end{array}
\right] \frac{1}{\vk_2(\eta,z)}
\\ \label{2.23} &
-\frac{\vk_1(\eta,z)}{\vk_2(\eta,z)^2}
[0 \quad 1]F(0,\eta,z)R(\eta,z)
\left[
\begin{array}{c}
\vp(0,z) \\1
\end{array}
\right].
\end{align}
Using the expression for $F(0,\eta,z)$ $($compare for that purpose \eqref{2.4} and \eqref{2.7'}$)$,
we rewrite \eqref{2.23} in the final form
\begin{align}&\nonumber
 \frac{d}{d \eta}\vp(\eta,z)=\frac{1}{z+\g}\Big(i\big(\sin 2h_2(\eta)\big)\big(\exp{2id(\eta)}\big)\vp(\eta,z)^2 
\\ \label{2.34} &
+  2\big(\cos 2h_2(\eta)\big)\vp(\eta,z)
+i\big(\sin 2h_2(\eta)\big)\big(\exp{-2id(\eta)}\big)\Big).
\end{align}
 \end{Rk}
Another case of Riccati equations for Weyl functions one can find, for instance, in \cite{GesZ}.
\section{Sine-Gordon equation in a semi-strip} \label{SGE}
\setcounter{equation}{0}
\subsection{Existence theorems and construction of  solution}
Theorem 3 in \cite{SaA2} gives sufficient conditions, under which
a solution of   sine-Gordon equation in the semi-strip ${\cal D}$ exists,
and a procedure to recover this solution. The procedure to solve the
sine-Gordon equation in the semi-strip is based on
the procedure to solve the inverse problem, which is given in
Theorem \ref{TmR}.
\begin{Dn}\label{DnOm} Let $\vp(z)$ be holomorphic in the semi-plane $\Im z<-M$
and admit  representation \eqref{1.4!}. Then, according to Theorems \ref{TmInv2} and \ref{TmR}
there is a unique solution of the inverse problem, that is, there is a unique potential
$v$ such that \eqref{1.2'} holds and $\vp$ is the Weyl function of the corresponding system 
\eqref{1.1}. Denote this solution of the inverse problem by $\Om(\vp)$  $($i.e., $v(\xi)=\big(\Om(\vp)\big)(\xi))$.
\end{Dn}
\begin{Tm}\label{Wsg}\cite{SaA2} Let the initial--boundary  conditions
\begin{equation} \label{3.1}
\psi(\xi,0)=h_1(\xi), \quad \psi(0,\eta)=h_2(\eta), \quad (h_1(0)=h_2(0), \quad h_k=\ov{h_k})
\end{equation}
be given. Assume that  $h_2$ is continuous on $[0, \, a)$ and that
$h_1$ is boundedly  differentiable on all the finite intervals on $[0,\, \infty)$.
Moreover, assume that the Weyl function $\vp_0(z)$ of the system
\begin{equation} \label{3.2}
W_{\xi}=GW, \quad G(\xi,z)=izj+V(\xi), \quad V(\xi)=\left[
\begin{array}{lr}
0& -h_1^{\prime}(\xi) \\  h_1^{\prime}(\xi) & 0
\end{array}
\right]
\end{equation}
exists
 and admits representation  \eqref{1.4!}. Then a solution of the
initial--boundary value problem \eqref{0.1}, \eqref{3.1}  exists
and is given by the equality
\begin{equation} \label{3.4}
\psi(\xi,\eta)=h_2(\eta)-\int_0^{\xi}\Big(\Om\big(\vp(\eta,z)\big)\Big)(x)dx,
\end{equation}
where
\begin{equation} \label{3.5}
\vp(\eta,z)=\frac{R_{11}(\eta,z)\vp_0(z)+R_{12}(\eta,z)}
{R_{21}(\eta,z)\vp_0(z)+R_{22}(\eta,z)}, 
\end{equation}
and $R=\{R_{kp}\}_{k,p=1}^2$ is defined by the relations
\begin{equation} \label{3.6}
\frac{d}{d \eta}R(\eta,z)=\frac{1}{i z}
\left[
\begin{array}{lr}
\cos 2h_2(\eta)&  \sin 2 h_2(\eta) \\ \sin 2 h_2(\eta) &-\cos 2 h_2(\eta)
\end{array}
\right]R(\eta,z), \quad R(0,z)=I_2.
\end{equation}
Here $\vp(\eta,z)$ admits representation \eqref{1.4!}, where
\[
\a(\eta)=\a(0)-i\int_0^{\eta}\sin\big(2h_2(x)\big)dx, 
\]
and we can put  $M(\eta)\equiv \wt M(\wt a)$ 
for all $\eta$ on the intervals  $ 0\leq \eta \leq \wt a$
for each $0<\wt a<a$ and some $\wt M(\wt a)>0$.
So $\Om$ is well-defined.
\end{Tm}
\begin{Rk} The equality
\begin{equation} \label{3.3}
\vp(\eta, z)=\ov{\vp(\eta,-\ov z)},
\end{equation}
where $\vp$ is given by \eqref{3.5},
is used in the proof of Theorem \ref{Wsg}  to show that $\psi=\ov{\psi}$.
It is also of independent interest. To derive \eqref{3.3} notice that 
the fundamental solution $w$ of  \eqref{1.1}, where $V$ is given in \eqref{3.2},
and the fundamental solution $R(\eta,z)$ of \eqref{3.6}
have the properties
\begin{equation} \label{3.3!}
w(\xi,  z)=\ov{w(\xi, -\ov z)}, \quad  R( \eta, z)=\ov{R( \eta,-\ov z)}.
\end{equation}
It follows from Definition \ref{Dn1} and \eqref{3.3!} that $\ov{\vp_0(-\ov z)}$
is the Weyl function of system \eqref{3.2} simultaneously with $\vp_0( z)$.
Hence, by Proposition \ref{P1} we have 
$\ov{\vp_0(-\ov z)}=\vp_0( z)$. Therefore, equality  \eqref{3.3}
is immediate from \eqref{3.5} and the second equality in \eqref{3.3!}.
\end{Rk}
The next proposition is a particular case of Theorem 6.1 in \cite{BC}
\begin{Pn} Suppose $v(\xi)\in L^1(\BR_+)$. Then there is a fundamental solution
$W$ of \eqref{1.1} such that  we have
\begin{equation} \label{3.7}
\lim_{z \to \infty}W(\xi,z)e^{-i\xi z j}=I_2, \quad z\in \BC_-
\end{equation}
uniformly with respect to $\xi$.
If, in addition, $v$ is two times differentiable 
and $v^{\prime}(\xi), \, v^{\prime \prime}(\xi)
\in L^1(\BR_+)$, then there is a matrix $E$ such that 
\begin{equation} \label{3.8}
W(0,z)=I_2+\frac{1}{z}E+O\big(\frac{1}{z^2}\big), \quad z\to \infty, \quad z\in\BC_-.
\end{equation}
\end{Pn}
\begin{Cy} \label{MCy} (i) If  $v \in L^1(\BR_+)$ then there is a Weyl function
of system \eqref{1.1} and this Weyl function is given by the formula
\begin{equation} \label{3.9}
\vp(z)=W_{12}(0,z)/W_{22}(0,z), \quad W(0,z)=:\{W_{kp}(0,z)\}_{k,p=1}^2
\end{equation}
for all $z$ in the semi-plane $\Im z<-M$ for some $M>0$.

(ii) If  $v$ is two times differentiable and $v, \, v^{\prime}, \, v^{\prime \prime}
\in L^1(\BR_+)$, then this Weyl function $\vp$ admits representation \eqref{1.4!}.
\end{Cy}
\begin{proof}. By \eqref{3.7} (see also Theorem A in \cite{BC})
there is a value $M>0$ such that $W(\xi,z)$
is holomorphic for $\Im z<-M$, and 
we have 
\begin{align} &\label{3.10}
\det W(0,z)\not=0, \quad W_{22}(0,z)\not=0, \\&\label{3.10!}
 \sup_{\xi \geq 0, \, \Im z<-M}\|W(\xi,z)e^{-i\xi z j}\|<\infty, \quad
\sup_{\Im z<-M}|1/W_{22}(0,z)|<\infty.
\end{align}
Thus, according  to \eqref{1.1}  it is immediate that 
\[
w(\xi,z)=W(\xi,z)W(0,z)^{-1}. 
\]
Hence, we derive
\begin{align}\nonumber &
e^{i \xi z} w(\xi, z)
\left[\begin{array}{c}
\frac{W_{12}(0,z)}{W_{22}(0,z)} \\ 1
\end{array}
\right]=\frac{e^{i \xi z}}{W_{22}(0,z)}W(\xi,z)W(0,z)^{-1}\left[\begin{array}{c}
W_{12}(0,z)\\ W_{22}(0,z) 
\end{array}
\right] \\  \label{3.11}&
=\frac{1}{W_{22}(0,z)}W(\xi,z)e^{-i\xi z j}\left[\begin{array}{c}
0\\ 1
\end{array}
\right].
\end{align}
In view of  \eqref{3.10!} and \eqref{3.11} the function $\vp$ given by
\eqref{3.9} satisfies conditions of Definition \ref{Dn1}, that is, the statement (i)
is proved.

If $v, \, v^{\prime}, \, v^{\prime \prime}
\in L^1(\BR_+)$, then it follows from \eqref{3.8} that
\[
W_{12}(0,z)=\frac{1}{z}E_{12}+O\big(\frac{1}{z^2}\big), \quad
W_{22}(0,z)=1 +\frac{1}{z}E_{22}+O\big(\frac{1}{z^2}\big).
\]
Therefore, we get for $z\in \BC_-$, $z \to \infty$ that
\begin{equation} \label{3.12}
1/W_{22}(0,z)=1-\frac{1}{z}E_{22}+O\big(\frac{1}{z^2}\big), \,\,
W_{12}(0,z)/W_{22}(0,z)=\frac{1}{z}E_{12}+O\big(\frac{1}{z^2}\big).
\end{equation}
The statement  (ii) is immediate from \eqref{3.9} and \eqref{3.12}.
\end{proof}
The next theorem easily follows from Theorem \ref{Wsg} and Corollary \ref{MCy}.
\begin{Tm}\label{Exist}  Assume  that
$h_1(\xi)=\ov{h_1(\xi)}$ is three times  differentiable for $\xi \geq 0$, that 
\[h_1^{\prime}, \, h_1^{\prime \prime}, \, 
h_1^{\prime \prime \prime}
\in L^1(\BR_+),
\]
and that  $h_2=\ov{h_2}$ is continuous on $[0, \, a)\,$ $\big(h_1(0)=h_2(0)\big)$.
Then the Weyl function $\vp_0(z)$ of the system \eqref{3.2}
exists
 and admits representation  \eqref{1.4!}. A solution of the
initial--boundary value problem  \eqref{3.1} for sine-Gordon equation  
\eqref{0.1} exists
and is given by the equalities \eqref{3.4} and \eqref{3.5},
where $R=\{R_{kp}\}_{k,p=1}^2$ is defined by the relations \eqref{3.6}.
\end{Tm}
Note that a rapid decay of $h_1^{\prime}$ was required in \cite{PLV}.
\begin{Rk}\label{cont} If the conditions of Theorem \ref{Exist} hold,
then by Lemma 2 from \cite{SaA2} the functions $\psi$, $\psi_{\xi}$,
and $\psi_{\xi \eta}$ are continuous.
 \end{Rk}

\subsection{Unbounded  solutions in the quarter-plane}
The behavior of the solutions of initial-boundary value problems is of
interest. Notice also that it is difficult to treat  unbounded solutions
using the Inverse Scattering Transform method. Here we describe
a family of unbounded solutions.

First we formulate Theorem 2 from \cite{SaA2}, which is proved in \cite{SaA2} in a way
quite similar to the proof of   Theorem \ref{evol}. (The equivalence of the definitions
of  Weyl functions here and in \cite{SaA2} follows  from Proposition \ref{P2}
under condition \eqref{1.3}.)
\begin{Tm}\label{evolSG}
Let $\psi=\ov{\psi}$ and $\psi_{\xi}$ be continuous functions in the semi-strip
${\mathcal{D}}$, let $\psi_{\xi \eta}$ exist, and let \eqref{0.1} hold. Assume that
$\psi_{\xi}$ is bounded, that is,
\begin{equation} \label{3.14}
\sup|\psi_{\xi}(\xi,\eta)|\leq M \quad \big( (\xi,\eta)\in {\mathcal{D}}\big).
\end{equation}
Then the evolution $\vp(\eta,z)$ of the Weyl function of the auxiliary system
$W_{\xi}=GW$, where $G$ is given by \eqref{0.13} and \eqref{0.14},
is expressed by the formula \eqref{3.5}, where $R(\eta,z)$ is given by \eqref{3.6},
$\, h_2(\eta)=\psi(0,\eta)$, and $\vp_0(z)=\vp(0,z)$.
\end{Tm}
Put $a=\infty$ in \eqref{0.9}, that is, let $\mathcal D$ be a quarter-plane.
Then the next corollary follows from Theorem \ref{evolSG}.
\begin{Cy}  \label{CyBVP} Assume that  $\,\,  \mathcal D \,\,$ is a quarter--plane
and the conditions of  Theorem \ref{evolSG} hold. Then, for values
of $z$, such that the inequalities
\begin{equation} \label{3.15}
\big(\cos \, 2h_2(\eta)-\ve(z) \big)\Im \ov{z}\geq |\sin \, 2h_2(\eta)||\Re z |, 
\quad \Im z<-M
\end{equation}
hold for some $\ve(z)>0$ and for all $\eta \geq0$, we have
\begin{equation} \label{3.16}
\vp_0(z)=-\lim_{\eta \to \infty}R_{12}(\eta,z)/R_{11}(\eta,z).
\end{equation}
\end{Cy}
\begin{proof}. Recall that according to Proposition \ref{P2} the
inequality 
\[
|\vp(\eta,z)| \leq 1
\]
 is true. Hence, in view of \eqref{3.5} we get
\begin{equation} \label{3.17}
[\vp_0(z)^* \quad 1]R(\eta,z)^*jR(\eta,z)\left[
\begin{array}{c}
\vp_0(z)  \\ 1
\end{array}
\right] \leq 0.
\end{equation}
By \eqref{3.6} and  \eqref{3.15} we derive
 \begin{equation} \label{3.18}
\frac{d}{d \eta}\Big(R(\eta,z)^*jR(\eta,z)\Big)
\geq  \frac{2\ve(z)}{|z|^2} \big(\Im \ov{z}\big) R(\eta,z)^*R(\eta,z).
\end{equation}
It is immediate from \eqref{3.17} and \eqref{3.18} that
\begin{equation} \label{3.19}
\int_0^{\infty}[\vp_0(z)^* \quad 1]R(\eta,z)^*R(\eta,z)\left[
\begin{array}{c}
\vp_0(z)  \\ 1
\end{array}
\right] d \eta <\infty.
\end{equation}
As according to \eqref{3.18} we have $\frac{d}{d \eta}\big(R(\eta,z)^*jR(\eta,z)\big)\geq 0$, it follows that
$R(\eta,z)^*jR(\eta,z)\geq j$. In particular, we get
 \begin{equation} \label{3.20}
|R_{11}(\eta, z)|\geq 1.
\end{equation}
According to \eqref{3.6} we have also $\frac{d}{d \eta}\Big((\exp 2\eta / |z|)R(\eta,z)^*R(\eta,z)\Big)\geq 0$,
that is,
 \begin{equation} \label{3.21}
R(\eta,z)^*R(\eta,z) \geq \big(\exp 2(\eta_0 - \eta) / |z|\big)R(\eta_0,z)^*R(\eta_0,z), \quad \eta \geq \eta_0\geq 0.
\end{equation}
Inequalities \eqref{3.19} and  \eqref{3.21} imply that
 \begin{equation} \label{3.22}
\lim_{\eta\to \infty}\left\| R(\eta,z)\left[
\begin{array}{c}
\vp_0(z)  \\ 1
\end{array}
\right]  \right\|=0.
\end{equation}
Finally, \eqref{3.16} follows from \eqref{3.20} and  \eqref{3.22}.
\end{proof}
\begin{Ee} \label{Unb}
Let $h_2(\eta)\equiv 0$ $(\infty>\eta\geq 0)$. Putting $\ve=1/2$ we see that \eqref{3.15} holds for all $z$ in the
semi-plane $\Im z<-M<0$. By \eqref{3.6} the equality
 \begin{equation} \label{3.23}
R(\eta, z)=e^{(\eta/iz)j}
\end{equation}
is true. Thus, if the conditions of Theorem \ref{evolSG} hold, by Corollary \ref{CyBVP} we
derive $\vp_0(z) \equiv 0.$
It is immediate from Definition \ref{Dn1}, that if  $\vp_0(z) \equiv 0$, then 
$v=\Om(\vp_0)\equiv 0$ (see also Definition \ref{DnOm} of  $\Om(\vp)$).
Therefore, taking into account that $\vp_0(z)=\vp(0,z)$, we have
\begin{equation} \label{3.24}
\psi_{\xi}(\xi,0)=-\Om(\vp_0)\equiv 0.
\end{equation}
\end{Ee}
From Example \ref{Unb} and Theorem \ref{Exist} we derive the next proposition.
\begin{Pn}\label{PnUnb} Assume  that
$h_1(\xi)=\ov{h_1(\xi)}\not\equiv 0$ is three times  differentiable for $\xi \geq 0$, that 
\[h_1^{\prime}, \, h_1^{\prime \prime}, \, 
h_1^{\prime \prime \prime}
\in L^1(\BR_+), \quad h_1(0)=0,
\]
and that  $h_2\equiv 0$. Then one can use
the procedure given in Theorem \ref{Exist} to construct a solution
$\psi$ of the initial-boundary value problem \eqref{3.1} for sine-Gordon
equation \eqref{0.1},
and  $\psi_{\xi}$ is always unbounded in the quarter-plane.
\end{Pn} 
\begin{proof}. As the conditions of Theorem \ref{Exist} hold, we can construct a solution $\psi$
of  \eqref{0.1}, \eqref{3.1}. Moreover, by Remark \ref{cont} the functions $\psi$, $\psi_{\xi}$, 
and $\psi_{\xi \eta}$ are
continuous. 

Thus, if \eqref{3.14} holds, then the conditions of Theorem \ref{evolSG}
and Example \ref{Unb} are satisfied. In particular, we get \eqref{3.24},
which contradicts the initial condition $\psi(\xi,0)=h_1(\xi)$ and assumptions 
\[
h_1(0)=0, \quad  h_1(\xi)\not\equiv 0
\]
of the proposition. So \eqref{3.14} is not true (i.e., the proposition is proved by contradiction).
 \end{proof}

{\bf Acknowledgement.}
The author is grateful to F. Gesztesy for fruitful discussions and for
his interest in the existence theorems, which stimulated this paper.
The work was supported by the Austrian Science Fund (FWF) under
Grant  no. Y330.


\begin{thebibliography}{AGKS}

\bibitem{Abd}
Abdellaoui, B., Colorado, E., and Peral, I.:
 Existence and nonexistence results for a class of linear and semilinear 
parabolic equations related to some 
Caffarelli-Kohn-Nirenberg inequalities.
 J. Eur. Math. Soc. (JEMS)  {\bf 6}:1,  119--148 (2004)

\bibitem{AKNS1}
Ablowitz, M. J.,  Kaup, D. J., Newell, A. C., and Segur, H.:  
Method for solving the sine-Gordon equation.
Phys. Rev. Lett.  {\bf 30}, 1262--1264  (1973) 

\bibitem{AKNS}
Ablowitz, M. J.,  Kaup, D. J., Newell, A. C., and Segur, H.: The inverse scattering transform -
Fourier analysis for nonlinear problems.  Stud. Appl. Math.  {\bf 53}:4,  249--315 (1974)

\bibitem{AS}
Ablowitz, M. J. and Segur, H.:  Solitons and the inverse scattering transform.
SIAM Studies in Applied Mathematics  {\bf 4}, Society for Industrial and Applied Mathematics (SIAM), 
Philadelphia, Pa., 1981.

\bibitem{AGKLS}
 Alpay, D., Gohberg, I., Kaashoek,  M.A., Lerer, L. and  Sakhnovich, A. L.:
Krein systems and canonical systems on a finite interval: accelerants with 
a jump discontinuity at the origin and continuous potentials.
arXiv:0912.4444 (2009)

\bibitem{Ba}
Bakas, I.: Conservation laws and geometry of perturbed coset models.  
Internat. J. Modern Phys. A  {\bf 9}:19,  3443--3472  (1994)

\bibitem{BP}
Barashenkov, I. V. and Pelinovsky, D. E.: Exact vortex solutions of the complex 
sine-Gordon theory on the plane.  Phys. Lett. B  {\bf 436}:1-2,  117--124   (1998)

\bibitem{BC}  Beals, R. and Coifman, R.R.:
 Scattering and inverse scattering for first-order systems.
Comm. Pure Appl. Math. {\bf 37}, 39--90 (1984)



\bibitem{Ber}
Berezanskii, Yu.M.: { Integration of non-linear  difference
equations by means of inverse problem technique}. Dokl. Akad. Nauk
SSSR {\bf  281}:1,  16--19 (1985)

\bibitem{BerG}
Berezanskii, Yu. M. and Gekhtman, M. I.:  Inverse problem of spectral analysis 
and nonabelian chains of nonlinear equations. 
Ukrainian Math. J.  {\bf  42}:6,  645--658  (1990)


\bibitem{BSZ} 
 Bona, J. L., Sun, S. M., and Zhang, B.-Yu.:  Boundary smoothing properties of the Korteweg-de Vries
equation in a quarter plane and applications. Dyn. Partial Differ. Equ.  {\bf 3}:1,  1--69 (2006)

\bibitem{Bona}
 Bona, J. L. and Winther, R.:  The Korteweg--de Vries equation, posed in a quarter-plane. SIAM J. Math. Anal. {\bf 14}, 1056--1106  (1983)


\bibitem{BoTz}
Bowcock, P. and  Tzamtzis, G.:  Quantum complex sine-Gordon model 
on a half line.   J. High Energy Phys.,    no. 11, 018, 22 pp  (2007)

\bibitem{CB}
Carroll, R. and Bu, Q.:
Solution of the forced nonlinear Schr\"odinger (NLS) equation using PDE techniques.
Appl. Anal. {\bf  41},  33--51 (1991)


\bibitem{CG2}  
Clark, F.   and  Gesztesy, F.: 
On Self-adjoint and J-self-adjoint Dirac-type Operators: A Case Study. Contemp. Math. {\bf 412}, 103--140 (2006)

\bibitem{DH} 
Dorey, N. and Hollowood, T.J.: 
Quantum scattering of charged solitons in the complex sine-Gordon model. 
Nuclear Phys. B {\bf  440}:1-2, 215--233  (1995)

 \bibitem{EY}
Escher,  J.  and Yin, Z.:
{ Well-posedness, blow-up phenomena, and global solutions for the $b$-equation}.
J. Reine Angew. Math. {\bf 624},  51-80 (2008)

\bibitem{TF}
Faddeev, L. D. and Takhtajan, L. A.:   Hamiltonian methods in the
theory of solitons. Springer, NY, 1986.


\bibitem{GesST}
Gesztesy, F., Simon, B., and Teschl, G., {Spectral deformations of one-dimensional 
Schr\"odinger operators}.   J. Anal. Math.  {\bf 70}, 267--324  (1996)

\bibitem{GesZ}
Gesztesy, F. and  Zinchenko, M.: {
Weyl-Titchmarsh theory for CMV operators associated with orthogonal polynomials 
on the unit circle}.   J. Approx. Theory  {\bf 139}:1-2,   172--213 (2006)

\bibitem{Hol}
Holmer, J.:  { The initial-boundary value problem for the Korteweg-de Vries equation}.
Comm. Partial Differential Equations  {\bf 31},  1151--1190  (2006)


\bibitem{IR}
Infeld, E. and  Rowlands, G.: { Nonlinear waves, solitons and chaos.} 
Second edition. Cambridge University Press, Cambridge, 2000.

\bibitem{KaS}
Kaashoek,  M. A. and Sakhnovich, A. L.:  {Discrete skew self-adjoint
canonical system and the isotropic Heisenberg magnet model}.  J.
Funct. Anal. {\bf 228}, 207--233 (2005)

\bibitem{KvM}
Kac,  M.  and van Moerbeke, P.:    { A complete solution of the
periodic Toda problem}. Proc. Natl. Acad. Sci. USA {\bf 72}, 2879--2880 (1975)

\bibitem{KN}
Kaup, D. J. and Newell, A. C.:  { The Goursat and Cauchy problems
for the sine-Gordon equation}.  SIAM J Appl. Math. {\bf 34}:1, 37--54 (1978)

\bibitem{Kri}
Krichever, I. M.: {
 An analogue of the d'Alembert formula for the equations of a principal chiral field 
and the sine-Gordon equation}. 
Soviet Math. Dokl. {\bf 22}:1,  79--84 (1980)

\bibitem{Lax}
Lax, P. D.: Integrals of nonlinear equations of evolution and solitary waves.  
Comm. Pure Appl. Math. {\bf 21}, 467--490  (1968)

\bibitem{LS}
Leznov, A. N. and Saveliev, M. V.: Group-theoretical methods for 
integration of nonlinear dynamical systems.  
Progress in Physics {\bf 15}, Birkh\"auser, Basel, 1992.

\bibitem{LR}
Lund, F. and  Regge, T.:
Unified approach to strings and vortices with soliton solutions.
Phys. Rev. D (3) {\bf 14}:6,  1524--1535  (1976)

\bibitem{MST}
Mennicken, R.,  Sakhnovich, A. L.,  and Tretter, C.: Direct and
inverse spectral problem for a system of differential equations
depending rationally on the spectral parameter.  Duke Math. J.
{\bf 109}:3, 413-449 (2001)

\bibitem{Nov}
Novikov, S. P.:
 A periodic problem for the Korteweg-de Vries equation.
Funct. Anal. Appl. {\bf 8}:3, 236--246 (1974)

\bibitem{PS1}
Park, Q-H., Shin, H. J.: Duality in complex sine-Gordon theory.  
Phys. Lett. B  {\bf 359}:1-2,  125--132  (1995)

\bibitem{PS2}
Park, Q-H., Shin, H. J.: Complex sine-Gordon equation in coherent optical
pulse propagation.
J. Korean Phys. Soc. {\bf 30}, 336-340 (1997)

\bibitem{Pol}
Pohlmeyer, K.:
Integrable Hamiltonian systems and interactions through quadratic constraints.
Comm. Math. Phys. {\bf 46}:3, 207--221  (1976)


\bibitem{SaA1}
Sakhnovich, A. L.:    Nonlinear Schr\"odinger equation on a
semi-axis and an inverse problem associated with it.  { Ukrainian Math . J.} {\bf 42}:3, 316--323  (1990)

\bibitem{SaA2}
Sakhnovich, A. L.:   The Goursat problem for the
sine-Gordon
equation and the inverse spectral problem.  { Russian
Math. (Iz. VUZ)} {\bf 36}:11, 42--52 (1992)

\bibitem{SaA3}
Sakhnovich, A.L.:   Spectral theory for  systems of
differential equations and applications. Thesis for
the secondary doctorship, Kiev, Institute of Mathematics, 1992.

\bibitem{SaA4}
Sakhnovich, A. L.: { Second harmonic generation: Goursat problem
on the semi-strip, Weyl functions and explicit solutions}. Inverse
Problems {\bf{21}}, 703-716 (2005)



\bibitem{SaA8}
{Sakhnovich},   A. L.: { Skew-self-adjoint discrete and
continuous
Dirac-type  systems:  inverse  problems and
Borg-Marchenko
theorems}. Inverse Problems {\bf 22}, 2083--2101 (2006)

\bibitem{SaA08}
Sakhnovich, A. L.: { Weyl functions, inverse problem and special solutions for the 
system auxiliary to the nonlinear optics equation}. Inverse Problems {\bf 24}, 025026  (2008) 

\bibitem{SaL20}
Sakhnovich, L. A.: { Non-linear equations and  inverse
problems on the semi-axis}. Preprint {\bf 30}, Inst. Mat. AN
Ukr.SSR, Izd-vo Inst. Matem. AN Ukr.SSR, Kiev, 1987.

\bibitem{Sa88} 
Sakhnovich, L. A.:  { Evolution of spectral data and nonlinear equations}. Ukrainian Math. J. {\bf 40}, 459--461 (1988)

\bibitem{SaL2} 
Sakhnovich, L. A.: The method of operator identities and problems in analysis.  
 St. Petersburg Math. J.  {\bf 5}:1, 1--69   (1994)


\bibitem{SaL30}
Sakhnovich, L. A.:  { Interpolation theory and its applications}.
Kluwer Academic Publishers, Dordrecht, 1997.

\bibitem{SaL3}
{Sakhnovich},   L .A.: { Spectral theory of canonical
differential
systems, method of operator identities}. Oper. Theory Adv. Appl. {\bf 107},  Birkh\"auser,  1999.

\bibitem{SCM}
Scott, A. C., Chu, F. Y. F., McLaughlin, D. W.:
 The soliton: a new concept in applied science. 
 Proc. IEEE  {\bf 61}, 1443--1483   (1973)
 

\bibitem{Ton} 
Ton, B. A.:
{  Initial boundary value problems for the Korteweg-de Vries equation}.
J. Differential Equations  {\bf 25}, 288-309   (1977)

\bibitem{PLV}
Vu, P.L.: 
The Dirichlet initial-boundary-value problems for sine and sinh-Gordon equations on a half-line.
Inverse Problems {\bf 21}:4, 1225-1248 (2005)

\bibitem{ZM}
Zakharov, V. E., Mikhailov, A. V.: Relativistically invariant two-dimensional models of field theory 
which are integrable by means of the inverse scattering problem method.  
Soviet Phys. JETP  {\bf 74}:6, 1953--1973  (1978)


\bibitem{ZTF}
Zakharov,  V.E., Takhtadzhyan,  L.A., Faddeev,  L.D.:
Complete description of solutions of the 'sine-Gordon' equation. 
Soviet Phys. Dokl. {\bf 19}, 824-826 (1974)
\end{thebibliography}
\end{document}